\documentclass[a4paper]{jpconf}
\usepackage[utf8x]{inputenc}
\usepackage[english]{babel}
\usepackage{amsmath}
\usepackage{multicol}
\usepackage{hyperref}
\usepackage{color}
\usepackage{graphicx}
\usepackage{epstopdf}

\begin{document}
\title{Finite size scaling in the dimer and six-vertex model}
  \author{P A Belov$^{1}$, A I Enin$^{1,2}$, A A Nazarov$^{1}$ }
  \address{$^{1}$Department of Physics, St. Petersburg State University, Ulyanovskaya 1, 198504 St.~Petersburg, Russia\\
    $^{2}$St. Petersburg Electrotechnical University, Professora Popova 5, 197376 St. Petersburg, Russia
  }
\ead{antonnaz@gmail.com}


\begin{abstract}
We present results of the Monte-Carlo simulations for scaling of the free energy in
dimers on the hexagonal lattice. The traditional Markov-chain Metropolis algorithm and more novel non-Markov Wang-Landau algorithm are applied. We compare the calculated results with the theoretical prediction for the equilateral hexagon and show that the latter algorithm gives more precise results for the dimer model.
For a non-hexagonal domain the theoretical results are not available, so we present the numerical results for a certain geometry of the domain. We also study the two-point correlation function in simulations of dimers and the six-vertex model. The logarithmic dependence of the correlation function on the distance, which is in accordance with the Gaussian free field description of fluctuations, is obtained.
\end{abstract}

\section{Introduction}

The six vertex model was introduced as the model of two-dimensional ice on a square lattice~\cite{lieb1967exact}. The oxygen atoms occupy the vertices of the lattice, the hydrogen atoms can be located on the edges close the the oxygen atom, but there can be only one hydrogen atom on each edge. The position of the hydrogen atom on the edge is then indicated by an arrow leading to the oxygen atom of the molecule. Thus there are exactly two incoming and two outgoing arrows at each vertex. The model was proposed as a toy model to study ice melting and crystallization, then became an important example of strong dependence of the scaling behavior on the boundary conditions \cite{korepin1982,tavares2015influence} and a test-bed for the transfer matrix method for the exact solution of lattice models \cite{baxter2016exactly}.

The dimer model appeared as  a very simplified model of solutions \cite{Fowler-1937}. The molecules are represented by the rigid tiles on a lattice. The partition function was computed on various lattices and with different boundary conditions \cite{doi:10.1080/14786436108243366, P.W-1961,kenyon2009lectures}. 

The six vertex and dimer models are the integrable lattice models of statistical physics. Since they allow exact solutions only for a few particular cases~\cite{baxter2016exactly,kulish1990yang}, both the models and their generalizations are under an active theoretical~\cite{zj2000,ferrari,mangazeev} and numerical investigation~\cite{allison2005numerical,ks2018}.

Specifically, the well-known limit shape phenomenon~\cite{vershik1977kerov} was discovered for these models~\cite{kenyon2006dimers} and a connection with the theory of random matrices was established~\cite{johansson2002non}.

For the studied models, the height function can be defined under some conditions. In the scaling limit, this function is fixed (``frozen'') outside of an analytical curve which is usually called ``the Arctic circle''~\cite{pronko2010}, though its shape depends on the particular model. Inside this curve, the height function weakly
converges to a certain surface. The fluctuations around this surface are described by the Gaussian free field.
Although the limit shape phenomenon~\cite{kenyon2006dimers} is studied by the advanced methods of algebraic geometry, the analytical description of the limit curve is obtained only for a few particular cases, such as domino tiling of the so-called ``Aztec diamond''~\cite{johansson2002non,kenyon2009lectures}.
From the point of view of the probability theory, limit shapes in the dimers and the six-vertex model are related to the Tracy-Widom distribution~\cite{tracy1994fredholm} of the random matrix eigenvalues.

Scaling behavior of the free energy depends on the geometry of the model and the area of the unfrozen domain. In this report, we present results of the Monte-Carlo simulations for scaling of the free energy in
dimers on the hexagonal lattice. The traditional Markov-chain Metropolis algorithm~\cite{metropolis1953equation} and more novel non-Markov Wang-Landau algorithm~\cite{wang2001efficient,landau2014guide} are applied for the study. We compare the calculated results with the theoretical prediction for the equilateral hexagon and show that the latter algorithm gives more precise results for the dimer model.
For a non-hexagonal domain the theoretical results are not available, so we present the numerical results for a certain geometry of the domain.

We also study the two-point correlation function in simulations of dimers and the six-vertex model.
We observe the appearance of the limit shape.
The logarithmic dependence of the correlation function on the distance, which is in accordance with the Gaussian free field description of fluctuations, is obtained.

\section{Lattice models}
\label{sec:dimers}

\subsection{The dimer model}

The configurations of the dimer model are perfect matchings of a bipartite graph with some choice of weights.
We consider coverings of the hexagonal lattice consisting of the subsets of lattice edges such that every vertex is the endpoint of exactly one edge.

\begin{figure}[htbp]
\center{\scalebox{0.4}{\includegraphics{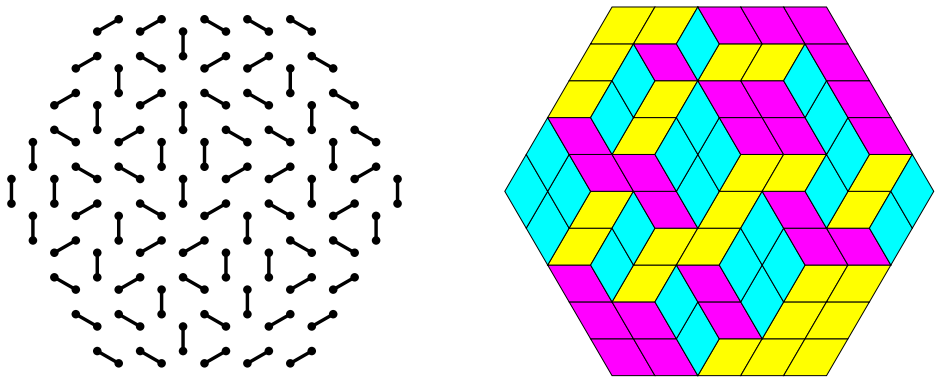}}}
  \caption{\label{dhf}A configuration of dimers on the hexagonal lattice and a corresponding picture of ``cubes in the corner''.}
\end{figure}

We can draw a rhombus on a dual lattice around each edge in the configuration. The picture of ``cubes in the corner'' presented in the Fig.~\ref{dhf} is obtained. Let us write on the top of each uppermost cube the height of its column of cubes. Looking at this picture from the top, we obtain a height function defined on the rectangular domain of the square lattice.

Let us define the sizes $M$, $N$, and $K$ of the sides of the hexagon.
The above description can be formalized by
setting the non-negative numbers up to $K$ in the boxes of the rectangular $M\times N$ table so that a value in
each box is not greater than values in the adjacent upper and left boxes
\begin{equation}
  \label{eq:1}
  h_{ij}\leq h_{i-1,j},\quad h_{ij}\leq h_{i,j-1}.
\end{equation}

The weight of a particular configuration is given by the exponent of the volume of all cubes or by a sum of the height function values:
\begin{equation*}
  \label{eq:18}
  E[conf]=\sum_{i,j} h_{ij}
\end{equation*}
We set Boltzman constant equal to 1, then the partition function is 
\begin{equation*}
  \label{eq:19}
  Z=\sum_{conf} e^{-\frac{E[conf]}{T}}=\sum_{conf}q^{\mathrm{Vol}[conf]}, 
\end{equation*}
where $q=\exp\left(-1/T\right)$ .

In general the partition function of the dimer model is given by the determinant of the Kasteleyn matrix \cite{P.W-1961,kenyon2009lectures,doi:10.1080/14786436108243366}, that can be seen as a discrete Dirac operator on a lattice \cite{kenyon2002laplacian}. 
For this particular case, the partition function is given by the classical Macmahon combinatorial formula~\cite{vuletic2009generalization}
\begin{equation}
  \label{eq:3}
   Z[M,N,K,q]=\prod_{i=1}^{M}\prod_{j=1}^{N}\prod_{k=1}^{K}\frac{1-q^{i+j+k-1}}{1-q^{i+j+k-2}}
\end{equation}

The scaling limit is achieved on the infinite lattice as $T\to\infty$. To study the scaling behavior using Monte-Carlo simulations on the finite lattices we will consider temperatures proportional to the lattice size and, then, extrapolate them to infinity.

\subsection{The six vertex model}
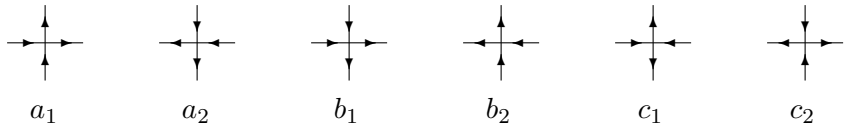
\begin{figure}[h]
\unitlength=1.0mm
\begin{center}
\begin{picture}(125,20)
\put(5,0){\line(0,1){10}}
\put(5,7.5){\vector(0,1){1}}
\put(5,2.5){\vector(0,1){1}}
\put(0,5){\line(1,0){10}}
\put(2.5,5){\vector(1,0){1}}
\put(7.5,5){\vector(1,0){1}}
\put(3,-5){$a_1$}
%
\put(25,0){\line(0,1){10}}
\put(25,7.5){\vector(0,-1){1}}
\put(25,2.5){\vector(0,-1){1}}
\put(20,5){\line(1,0){10}}
\put(22.5,5){\vector(-1,0){1}}
\put(27.5,5){\vector(-1,0){1}}
\put(23,-5){$a_2$}
%
\put(45,0){\line(0,1){10}}
\put(45,2.5){\vector(0,-1){1}}
\put(45,7.5){\vector(0,-1){1}}
\put(40,5){\line(1,0){10}}
\put(42.5,5){\vector(1,0){1}}
\put(47.5,5){\vector(1,0){1}}
\put(43,-5){$b_1$}
%
\put(65,0){\line(0,1){10}}
\put(65,2.5){\vector(0,1){1}}
\put(65,7.5){\vector(0,1){1}}
\put(60,5){\line(1,0){10}}
\put(62.5,5){\vector(-1,0){1}}
\put(67.5,5){\vector(-1,0){1}}
\put(63,-5){$b_2$}
%

\put(85,0){\line(0,1){10}}
\put(85,7.5){\vector(0,1){1}}
\put(85,2.5){\vector(0,-1){1}}
\put(80,5){\line(1,0){10}}
\put(82.5,5){\vector(1,0){1}}
\put(87.5,5){\vector(-1,0){1}}
\put(83,-5){$c_1$}
%
\put(105,0){\line(0,1){10}}
\put(105,7.5){\vector(0,-1){1}}
\put(105,2.5){\vector(0,1){1}}
\put(100,5){\line(1,0){10}}
\put(102.5,5){\vector(-1,0){1}}
\put(107.5,5){\vector(1,0){1}}
\put(103,-5){$c_2$}
\end{picture}
\vspace{5.mm}
\caption{The six types of vertices and the associated Boltzmann weights of the six vertex model.}
\label{6vert}
\end{center}
\end{figure}

The six vertex model is introduced as the model of two-dimensional ice on a square lattice~\cite{lieb1967exact,baxter2016exactly}.
Each edge of the lattice obtains an orientation, an arrow, in such a way that for each vertex there are exactly two incoming and two outgoing edges, so-called ``ice rule''.
The six types of possible vertices are presented in Fig.~\ref{6vert}.

The partition function of the six vertex model is given by a sum of all configurations:
\begin{equation*}
Z=\sum_{conf} \prod_{i=1}^M \prod_{j=1}^N \omega^{(i,j)}_{conf},
\label{partZ}
\end{equation*}
where the weight $w_{conf}^{(i,j)}$ for each vertex takes a predefine value according to Fig.~\ref{6vert}.
These Boltzmann weights are the matrix elements of the $R$-matrix, which is the key ingredient for the integrability and satisfy the Yang-Baxter equation~\cite{baxter2016exactly,kulish1990yang}.

The behavior of the six-vertex model strongly depends on the boundary conditions~\cite{korepin1982,tavares2015influence}.
We consider the model with the so-called ``domain wall boundary conditions''. Considering the domain as a vertical rectangle, it means that the oriented edges are entering the domain on the top and bottom borders and leaving the domain on the left and right borders.

We can represent the model with such conditions as a set of non-intersecting paths as follows. The paths start at the top border, can go along the arrows only downward or rightward. Each edge can belong to at most one path. The paths end on the right border. 
These paths are regarded as level lines of the height function. So we can represent the model in simulations by the rectangular table of height function values $h_{ij}$. 

The universality class of the scaling behavior is determined by the value of the combination
\begin{equation}
  \label{eq:21}
  \Delta=\frac{a^{2}+b^{2}-c^{2}}{2ab}.
\end{equation}
We study the case $\Delta=0$ when the limit shape phenomenon takes place and fluctuations are
described by the Gaussian free field~\cite{reshetikhin2016limit,reshetikhin2017integrability}.

\section{Simulation algorithms}
\label{sec:simul-algor}

\subsection{Metropolis algorithm}
The classical Metropolis algorithm based on the Markov chains can be applied to simulate both the dimer and the six vertex models~\cite{allison2005numerical,ks2018}.
The Markov chain is used to generate a random sequence of the model states.
The average of some observable over these states tends to the
expectation value
as a number of states grows.
The crucial requirement for the correctness of the algorithm is the detailed balance condition for the transition probabilities.

The following naive algorithm was implemented for the dimer model:
\begin{enumerate}
\item choose a random box $(i,j)$ in a table and change the height function value $\Delta h=\pm 1$ randomly;
\item if such a change does not break the condition~\eqref{eq:1} for the height function, we change the height by $\Delta h$ with the probability  $\exp\left(-{\Delta h}/{T}\right)$, otherwise we stay in the same configuration;
\item turn to the first step.
\end{enumerate}
Averaging over the configurations obtained on the second step of the algorithm, we obtain estimates for the thermodynamic observables.

Note that the detailed balance condition holds since the probability to choose the transition from a configuration to another one with one more cube is the same as the probability to choose a
transition to a configuration with one less cube.
The described algorithm can be optimized by saving a list of boxes where an addition or decrease of height function is allowed.
However, one must preserve the detailed balance condition and should choose a box on the first step with the probability equal to the ratio of possible changes in current and new configurations.
Note also that the Metropolis algorithm suffers from the critical slowdown. This problem is usually avoided by the use of Wolff~\cite{wolff1989collective} or Swendsen-Wang~\cite{swendsen1987nonuniversal} cluster algorithms.
In our case, we cannot apply cluster algorithms, since we need to preserve the condition~\eqref{eq:1} and similar one for the six vertex model. 

The free energy, $F=-T\log Z$, is unobtainable for any particular temperature because the Metropolis algorithm does not calculate the partition function, $Z$.
We have to use the relation
\begin{equation}
  \label{eq:9}
  F=\left< E \right>-TS
\end{equation}
The entropy $S$ can be obtained in simulations by a numerical integration of the heat capacity:
\begin{equation}
  \label{eq:10}
  S=\int_{0}^{T} \frac{C(t)}{t}dt, \quad C(T)=\frac{1}{T^{2}} \left(\left<E^{2}\right> -\left< E\right>^{2}\right).
\end{equation}
Therefore, we applied the Metropolis algorithm for various temperatures $T_{i}$ and obtained the values of observables $C(T_{i}), \left<E(T_{i})\right>$.
We calculated the entropy as a sum
$S(T_{i})=\sum_{j=1}^{i} C(T_{j}) (T_{j}-T_{j-1})/{T_{j}}$ and, as a result, computed a list of free energies $F(T_{i})$.

The Metropolis algorithm can be applied to the six vertex model \cite{allison2005numerical} in a similar way.

\subsection{The Wang-Landau algorithm}
We use the Wang-Landau algorithm~\cite{wang2001efficient,landau2014guide} to simulate the energy distribution $\rho(E)=e^{g(E)}$ of the dimer model.

The realization of this algorith is as follows~\cite{bns,bnsconf}.
The energy range is split into some number of intervals, which can coincide with the number of discrete energies.
The algorithm starts with a random configuration (random height function which satisfies the condition~\eqref{eq:1}), an empty array of logarithms of energy densities $g(E_{1}),\dots,g(E_{s})$, an empty visitation histogram $n(E_{1}),\dots,n(E_{s})$ and some initial value (usually 1) of constant $a$.
On each step of the algorithm, a box $(i,j)$ and a change of the value of the height function $\Delta h$ at this box are chosen randomly.
A new configuration with the changed energy value $h_{ij}+\Delta h$ is accepted with the probability $e^{g(E_{new})-g(E_{old})}$ if the condition \eqref{eq:1} is satisfied. The visitation number $n(E)$ is increased by 1 and $g(E)$ is increased by $a$. Otherwise, the current configuration is kept, $n(E)$ and $g(E)$ remain unchanged. This procedure is repeated until the visitation histogram becomes relatively flat, i.e. $\min(n(E))/\max(n(E))>0.8$. Then, the value of $a$ is divided by 2, the histogram is emptied and the next step of the algorithm begins.

In the limit, the distribution $\rho(E)$ becomes stationary. In practice, about 25 such steps are enough to do in simulations.
Having the energy distribution $\rho(E)$, the partition function is
given by
\begin{equation}
  \label{eq:8}
  Z=\sum_{E_{i}} \rho(E_{i}) e^{-\frac{E_{i}}{T}}.
\end{equation}

Calculation of the partition function is an important advantage of the Wang-Landau algorithm over the Metropolis~\cite{metropolis1953equation} and Wolff algorithms~\cite{wolff1989collective}.
Having the partition function, we obtain the free energy $F=-T\log Z$.

\begin{figure}[h!]
    \includegraphics[scale=0.6]{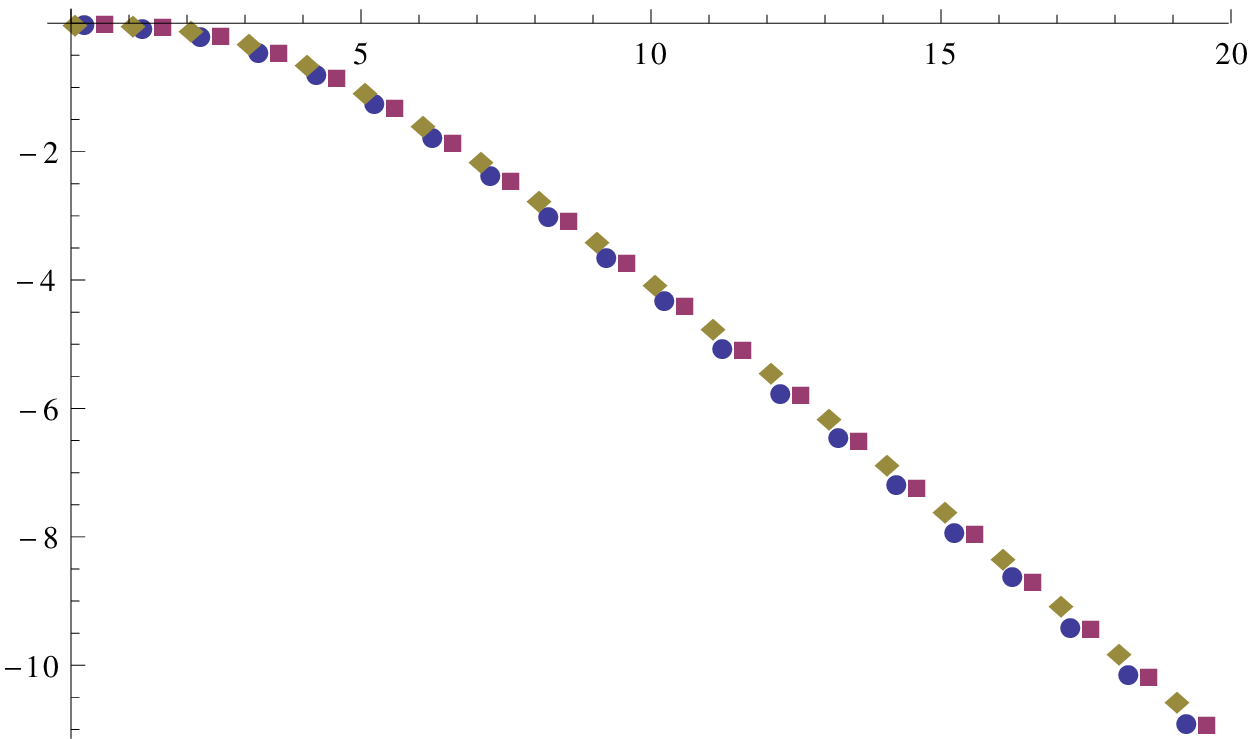}
  \includegraphics[scale=0.6]{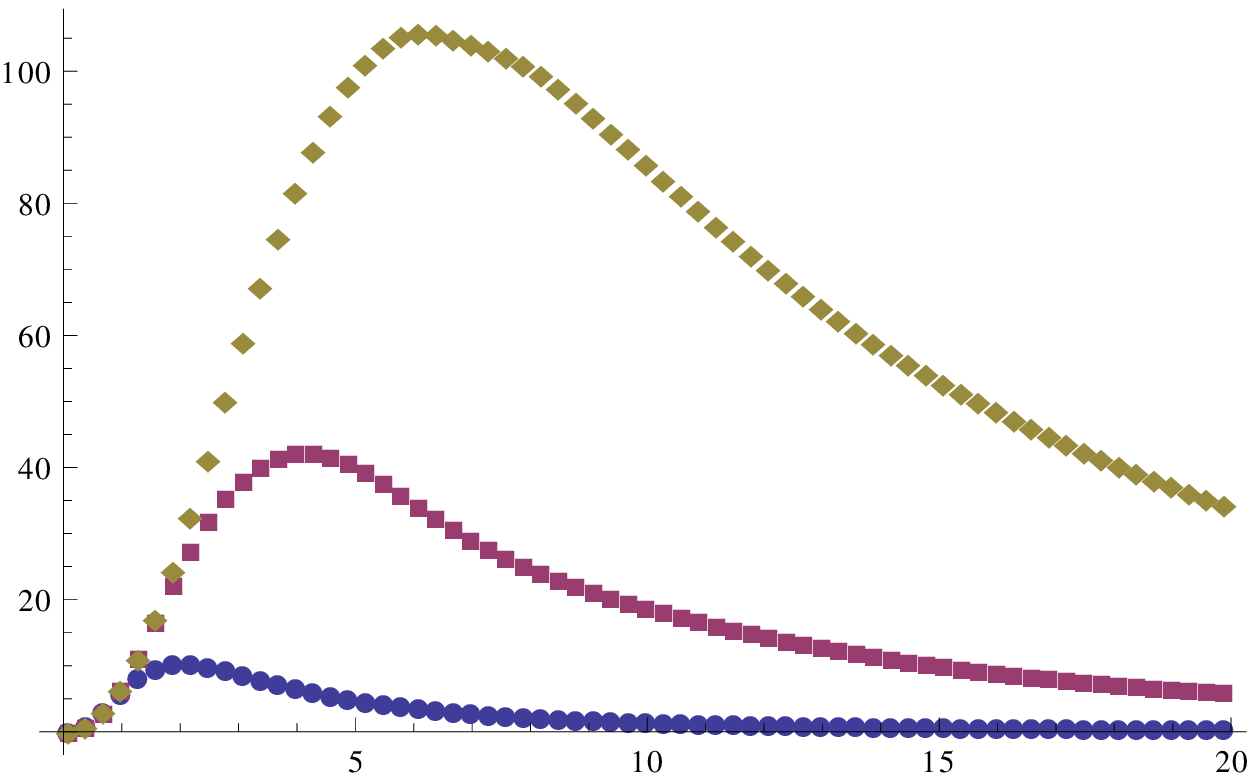}
  \label{fig:free-energy-10-10}
  \caption{\emph{Left panel:} A temperature dependence of the heat capacity for the dimer model on the equilateral hexagonal lattice: the Metropolis data (circles), the Wang-Landau data (diamonds), theoretical values (squares). \emph{Right panel:} A temperature dependence of the free energy for the dimer model on the equilateral hexagonal lattice, obtained from the Wang-Landau simulation for side legths: 5 (circles), 10 (squares), 16 (diamonds).}
\end{figure}

An example of simulation results for the dimer model on the equilateral hexagonal lattice with the side length equal to 10 is presented in Fig.~\ref{fig:free-energy-10-10}~(left panel).

\section{Free energy scaling in the dimer model}
\label{sec:free-energy-scaling}

Let us fix a domain $D$ and cover it by a lattice with the lattice constant $\varepsilon$. Compute the logarithm of the partition function for the dimer model and consider its dependence on $\varepsilon$.

We made preliminary computations, expanding formula \eqref{eq:3} in $\varepsilon$.  There are also results in the literature for scaling of the determinant of a discrete Laplace operator \cite{kenyon2000asymptotic,sridhar2015asymptotic} and partial results on Dirac operator (Kasteleyn matrix) \cite{kenyon2002laplacian}. 
From these results we conjecture the scaling behavior
\begin{equation}
  \label{eq:4}
  \ln (Z_{D,\varepsilon})=\frac{1}{\varepsilon^{2}}  f_{0}
  +\frac{1}{\varepsilon} f_{1} + f_{2} \ln \varepsilon +f_{3},
\end{equation}
where $f_{i}$ are the fitted parameters.
We will present our computations in a separate publication. Here we demonstrate the numerical evidence and simulation results that  support this conjecture. 

Consider the equilateral hexagon with a side $M$ as the domain, then $\varepsilon=\frac{1}{M}$.
The scaling limit takes place on an infinite lattice with $q=\exp\left(-1/T\right)$ tending to one, i.e. $T\to\infty$. We need to consider temperatures proportional to the lattice size, $T=\tau M$, where $\tau$ is a dimensionless factor. So in the extrapolation to infinity, we keep $\tau$ to be constant.
We divide the formula \eqref{eq:4} by the volume, $M^{2}$, and multiply by $-T=-\tau M$. Scaling behavior of the free energy density becomes as following:
\begin{equation}
  \label{eq:6}
  f(\tau, \varepsilon)=-\frac{T}{M^{2}}\ln (Z[M,M,M,T])=-\frac{\tau}{\varepsilon}  f_{0}  -\tau
  f_{1} + \tau \varepsilon f_{2} \ln(\varepsilon)  -\tau \varepsilon f_{3} 
 \end{equation}

\begin{figure}[htbp]
  \label{logz-theoretic-fit}
  \includegraphics[scale=0.6]{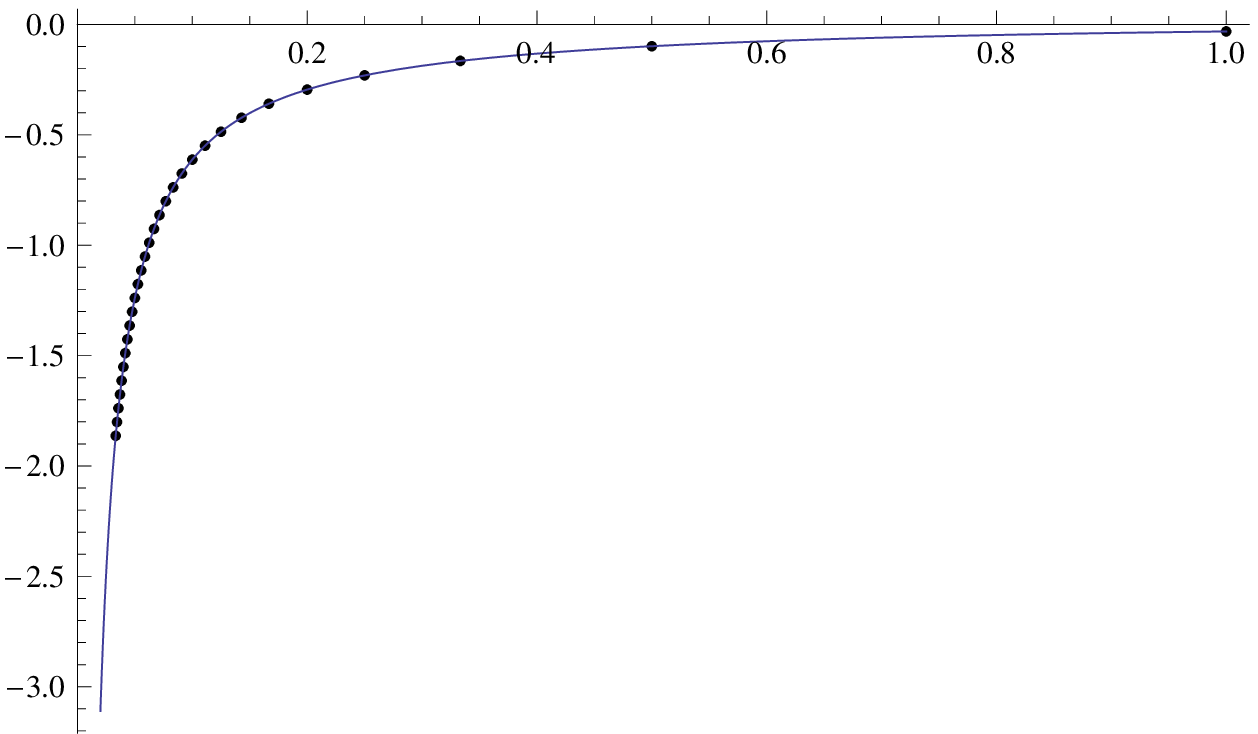}
  \includegraphics[scale=0.6]{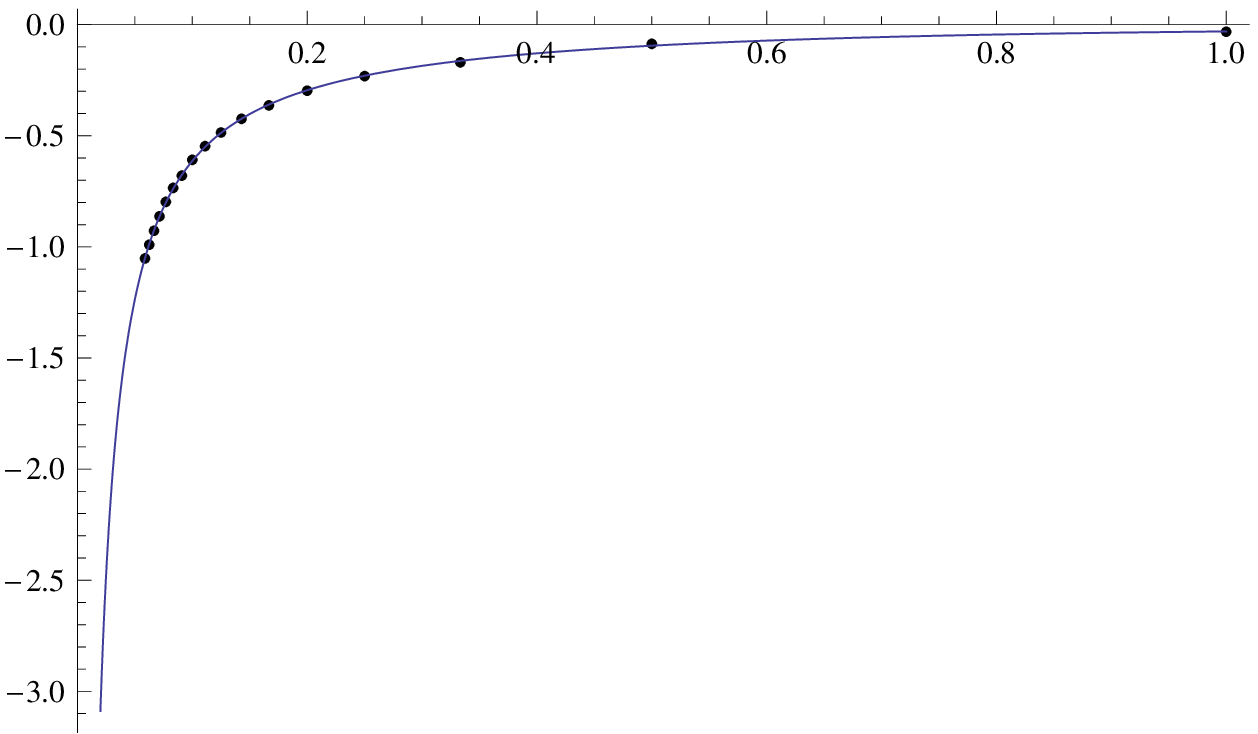}
  \caption{\emph{Left panel:} Theoretical dependence of the free energy density on the inverse lattice size $\varepsilon$ and fit by Eq.~\eqref{eq:6} for $\tau=\frac{2}{5}$. \emph{Right panel:} Results obtained with the Wang-Landau algorithm.}
\end{figure}

First, we test the expansion~\eqref{eq:4} by the Macmahon formula~\eqref{eq:3}.
Using it for lattice sizes ranging from 1 to 20 and doing a fit by Eq.~\eqref{eq:6}, we obtain the scaling parameters
\begin{equation}
  \label{eq:7}
  f_{0}=0.155704(5), \quad     f_{1}= -0.0019(2), \quad f_{2}=0.0741(4),\quad f_{3}=-0.0747(2)
\end{equation}
The uncertainty of the last decimal digit is given in brackets. The fit is shown in Fig.~\ref{logz-theoretic-fit} and the results are statistically significant for $f_{0}$, $f_{2}$, $f_{3}$. Therefore, our scaling behaviour, Eq.~\eqref{eq:4}, is supported by the numerical data.

Now we compare the results of Monte-Carlo simulations with the theoretical predictions~\eqref{eq:7}. We used the Wang-Landau algorithm with
25 iterations and the flatness to be 0.8 on equilateral hexagonal lattices with side ranging from 1 to 17. We
then made a fit by formula~\eqref{eq:6} and obtained the values
\begin{equation}
  \label{eq:13}
  f_{0}=0.154(2), \quad f_{1}= 0.04(4), \quad f_{2}=0.15(8),\quad f_{3}=-0.11(4)
\end{equation}
The uncertainty of the last decimal digit is given in brackets.
Although the calculated results are less precise, the Monte-Carlo results contain the theoretical predictions within the uncertainties. The corresponding fit of the free energy density is shown in the right panel of Fig.~\ref{logz-theoretic-fit}.

Using the Metropolis algorithm on the lattices of size from 3 to 18, we obtained similar but even less precise values
\begin{equation}
  \label{eq:12}
  f_{0}=0.15(1), \quad f_{1}= 0.2(3), \quad f_{2}=0(1),\quad f_{3}=0.0(4)
\end{equation}
This loss of precision is due to the critical slowdown that cannot be avoided since we are interested in the critical scaling behavior.

\begin{figure}[htbp]
  \label{corner-fit}
  \includegraphics[scale=0.6]{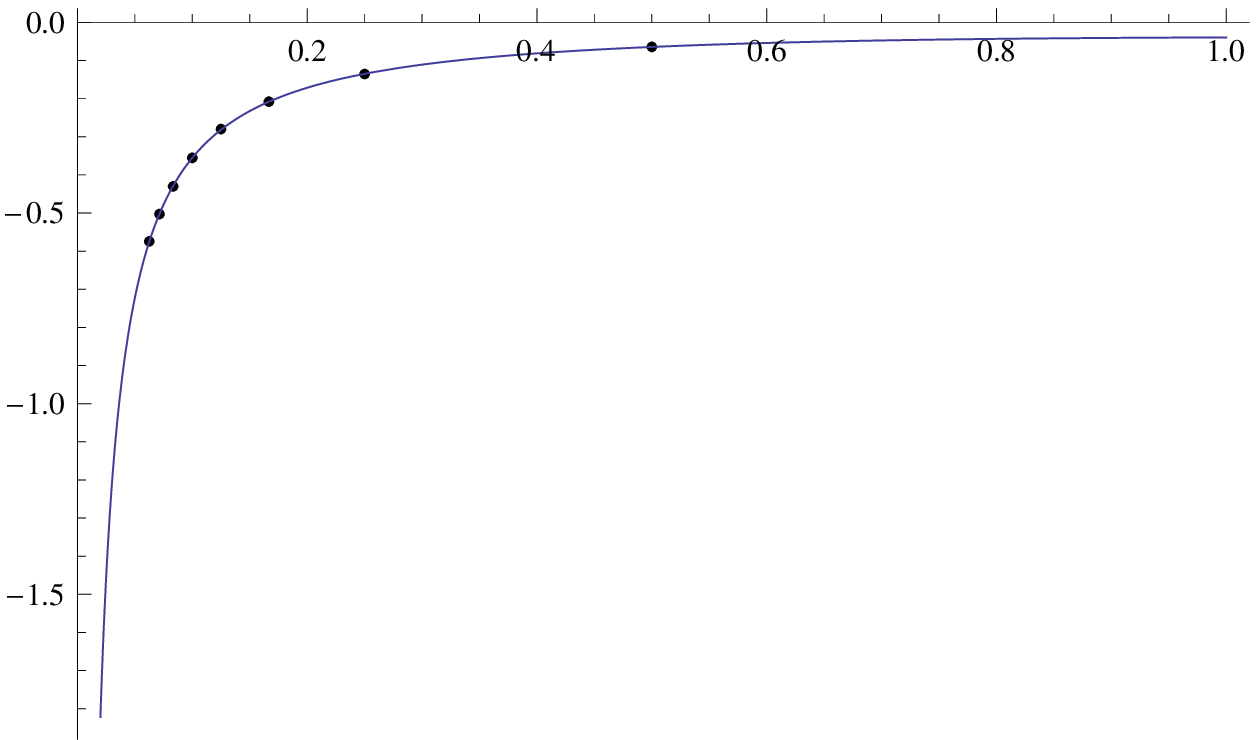}
  \includegraphics[scale=0.6]{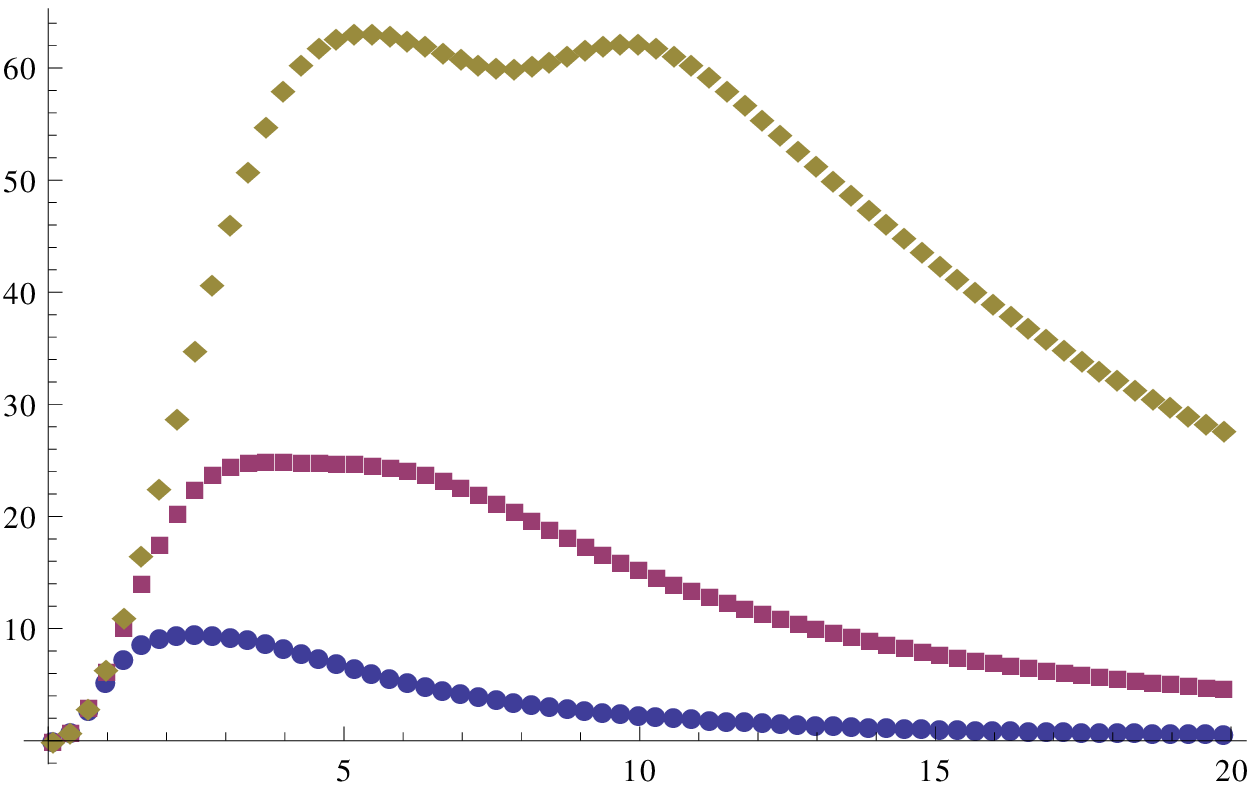}
  \caption{\emph{Left panel:} A dependence of the free energy density on the lattice constant $\varepsilon$ for the domain~\eqref{eq:5} from Wang-Landau simulations and fit by formula~\eqref{eq:6}. \emph{Right panel:} A the heat capacity dependence on the temperature for the domain~\eqref{eq:5} for $M=6$(circles), $M=10$ (squares) and $M=16$ (diamonds).}
\end{figure}

From the described comparisons, we see that the Wang-Landau algorithm is more efficient for study scaling behavior.
We can apply it to study the scaling behavior~\eqref{eq:4} on domains with more complex geometry, since there is no known formula for the partition function on such domains.
\begin{figure}[htbp]
  \label{corner-limit-shape}
  \includegraphics[scale=0.18,angle=90]{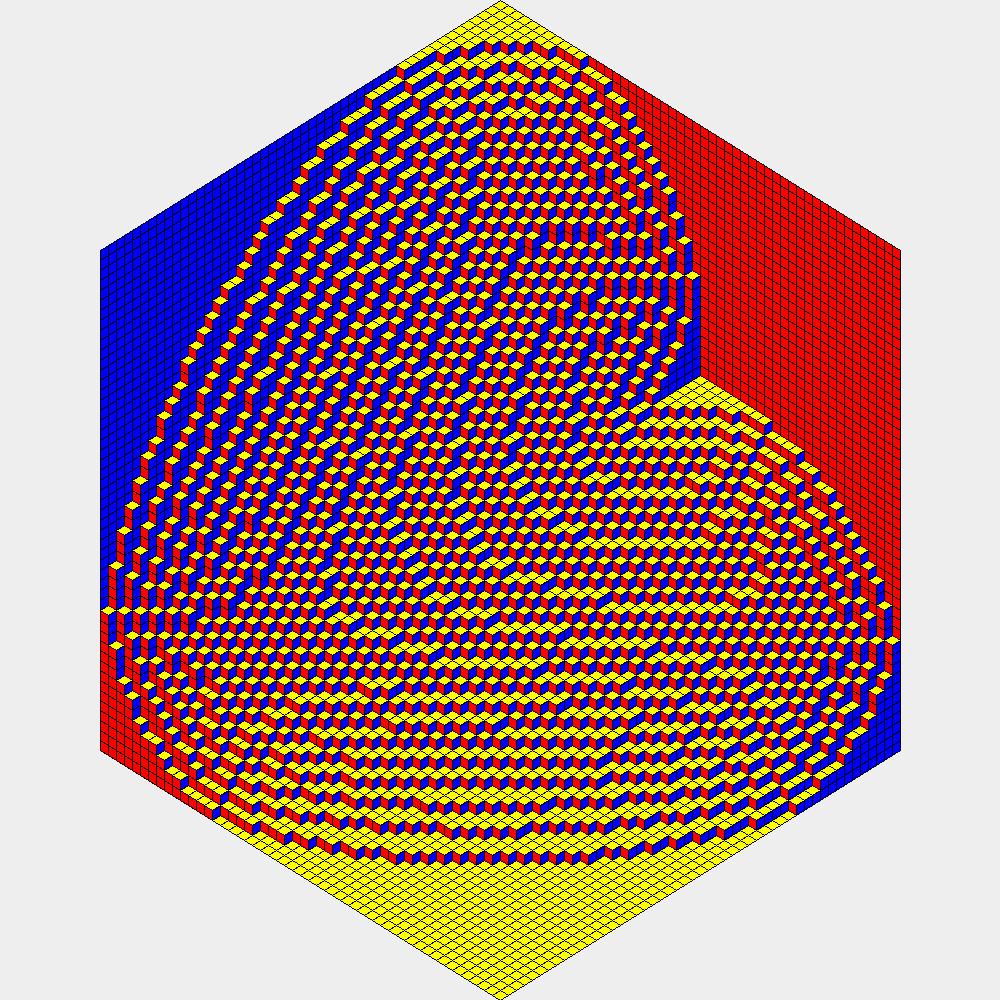}
  \includegraphics[scale=0.2]{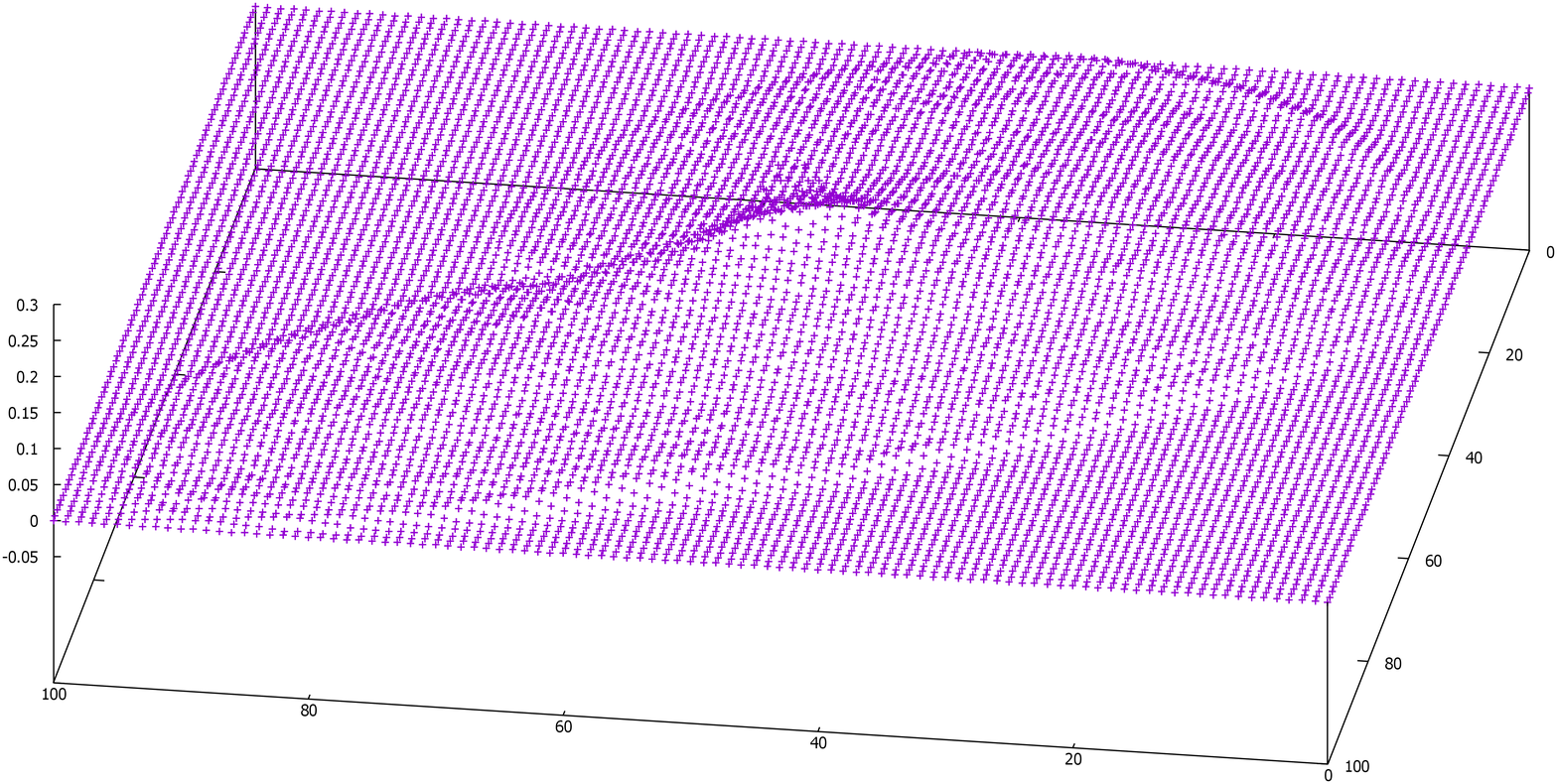}
  \caption{\emph{Left panel:} The limit shape for the domain~\eqref{eq:5} on the lattice of size $50\times 50$. \emph{Right panel:} the two-point height-height correlation function for the domain~\eqref{eq:5}.}
\end{figure}
For example, from Wang-Landau simulations we easily obtained the scaling parameters of the free energy if the height function $h_{ij}$
is restricted to the domain
\begin{equation}
  \label{eq:5}
    \begin{cases} 
   i>M/2 & \text{if } j<M/2 \\
   i<M   & \text{if } j>M/2
  \end{cases}
\end{equation}
So, the region $i<M/2$ for $j<M/2$ is forbidden. The limit shape and the two-point correlation function for a domain with a cutted square region is shown in Fig.~\ref{corner-limit-shape}.
A dependence of the free energy density on the lattice constant is presented in the left panel of Fig.~\ref{corner-fit}. On the right panel, we show a behavior of the heat capacity on the temperature.

Fitted values of the scaling parameters for the domain~\eqref{eq:5} are:
\begin{equation}
  \label{eq:11}
  f_{0}=0.105(2), \quad     f_{1}= -0.02(4), \quad f_{2}=0.0(1),\quad f_{3}=0.03(2)
\end{equation}
We see that $f_{0}$ strongly depends upon the domain geometry. To draw some conclusion about $f_{1}$, $f_{2}$, $f_{3}$, more
extensive simulations are required.

\section{Two-point correlation functions and the limit shape in the dimer model}
\label{sec:two-point-corr-dimers}

Scaling behavior of the dimer model  inside the  ``the Arctic circle'' is described by an effective free field theory -- Gaussian free field (massless free boson)~\cite{kenyon2001dominos}. In the scaling limit height function converges to a free bosonic field. Thus, two-point height-height correlation function $\left<h(x)h(y)\right>$ in the dimer model inside ``the Arctic circle'' demonstrates logarithmic behavior for $x\neq y$~\cite{kenyon2009lectures}:
\begin{equation}
\label{eq:11.1}
\left<h(x)h(y)\right> =-\frac{1}{2\pi} \log{|x-y|}.
\end{equation}
This logarithmic behavior corresponds to the liquid phase of the model. Phase diagram and limit shape phenomenon is discussed in the paper \cite{kenyon2006dimers}.

\begin{figure}[h!btp]
  \label{fig:dimer-two-point-corr}
  \hspace{-2cm}\includegraphics[scale=0.2]{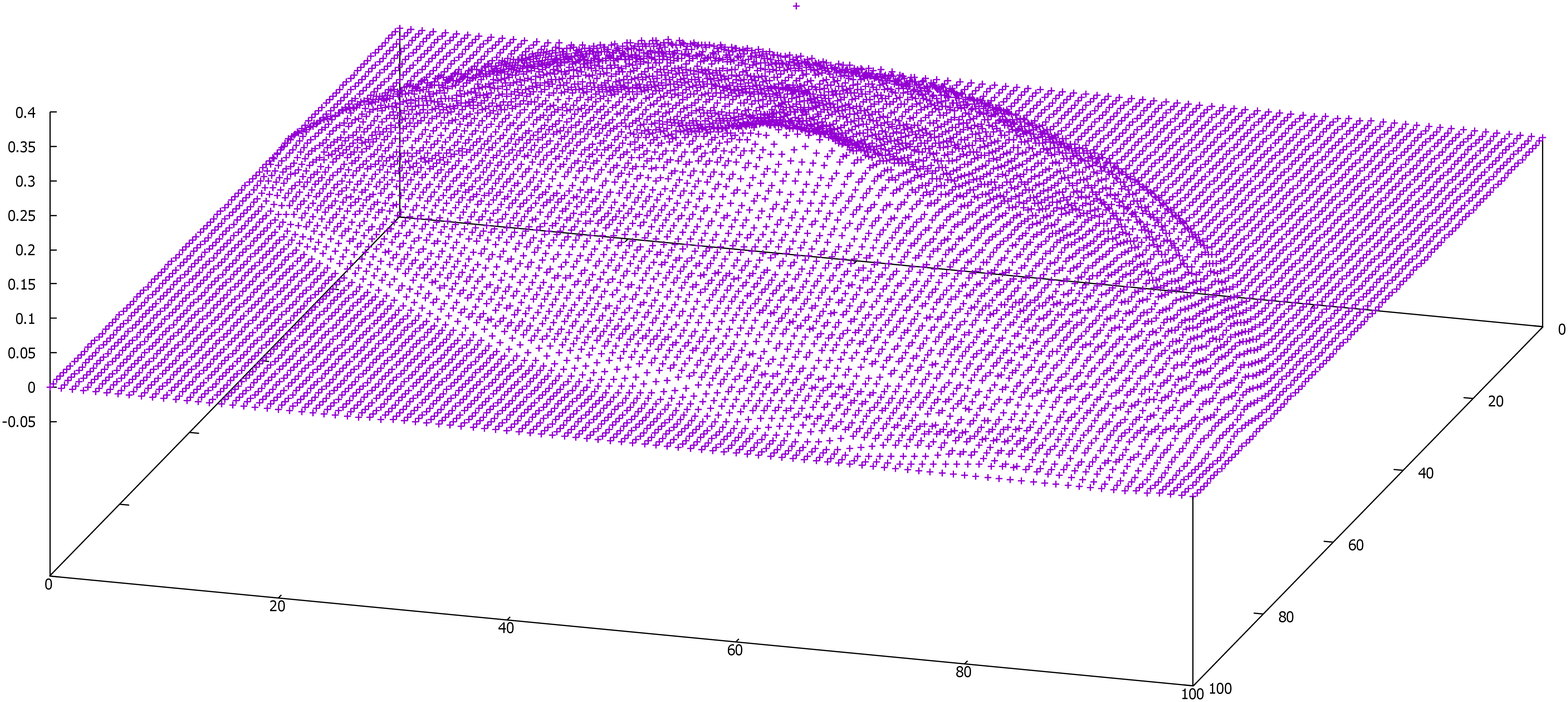}\hspace{-2cm}
  \includegraphics[scale=0.35]{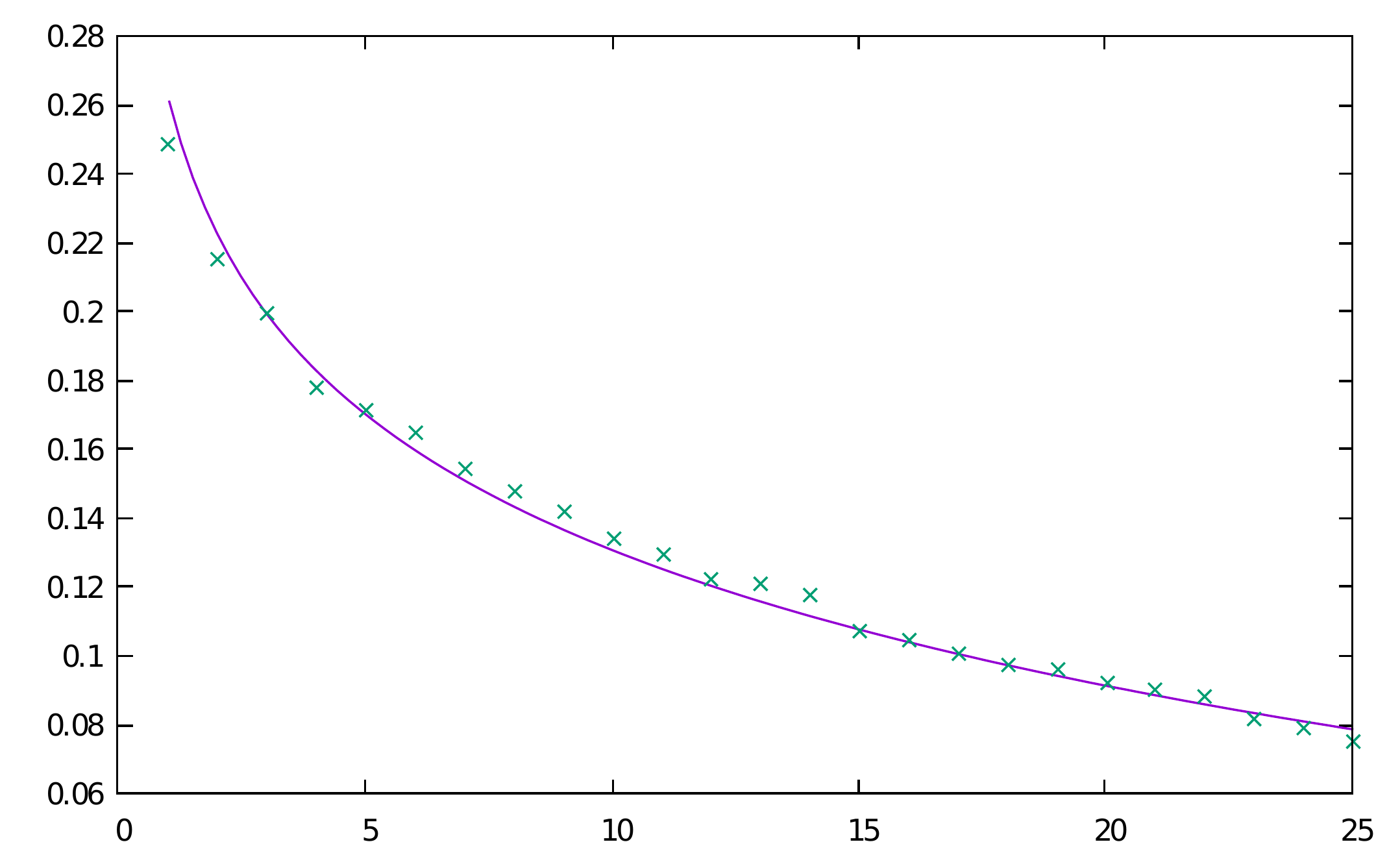}
  \caption{\emph{Left panel:} the two-point correlation function for the dimer model, the limit shape is clearly seen. The value $\left<h(y)^{2}\right>$ at the point $y=(50,50)$ in the center is used for the normalization of the correlation function. \emph{Right panel:} a dependence of the two-point correlation function on a distance and a logarithmic fit of the correlation function on $50\times 50$ lattice .}
\end{figure}

The simulated data is presented in Fig.~\ref{fig:dimer-two-point-corr}. 
Fixing, for example, $y$ and fitting the simulated data by a particular formula $a\log \left(\frac{x-y}{M}\right)$, as shown in Fig.~\ref{fig:dimer-two-point-corr} we obtain that $a=-0.0567(3)$. If we now choose the normalization of the correlation function in such a way that $\left<h(y)^{2}\right>=1$, we obtain the coefficient $-0.147(8)$ which is sufficiently close to $-\frac{1}{2\pi}$.
In the paper~\cite{kenyon2008height} it was proven that height fluctuations around the limit shape
in the scaling limit are described by the Gaussian free field. Our simulation result is in
agreement with this theoretical observation.
\section{Two-point correlation functions and the limit shape in the six vertex model}
\label{sec:two-point-corr-six-vertex}

\begin{figure}[htbp]
  \includegraphics[scale=0.43,angle=-90]{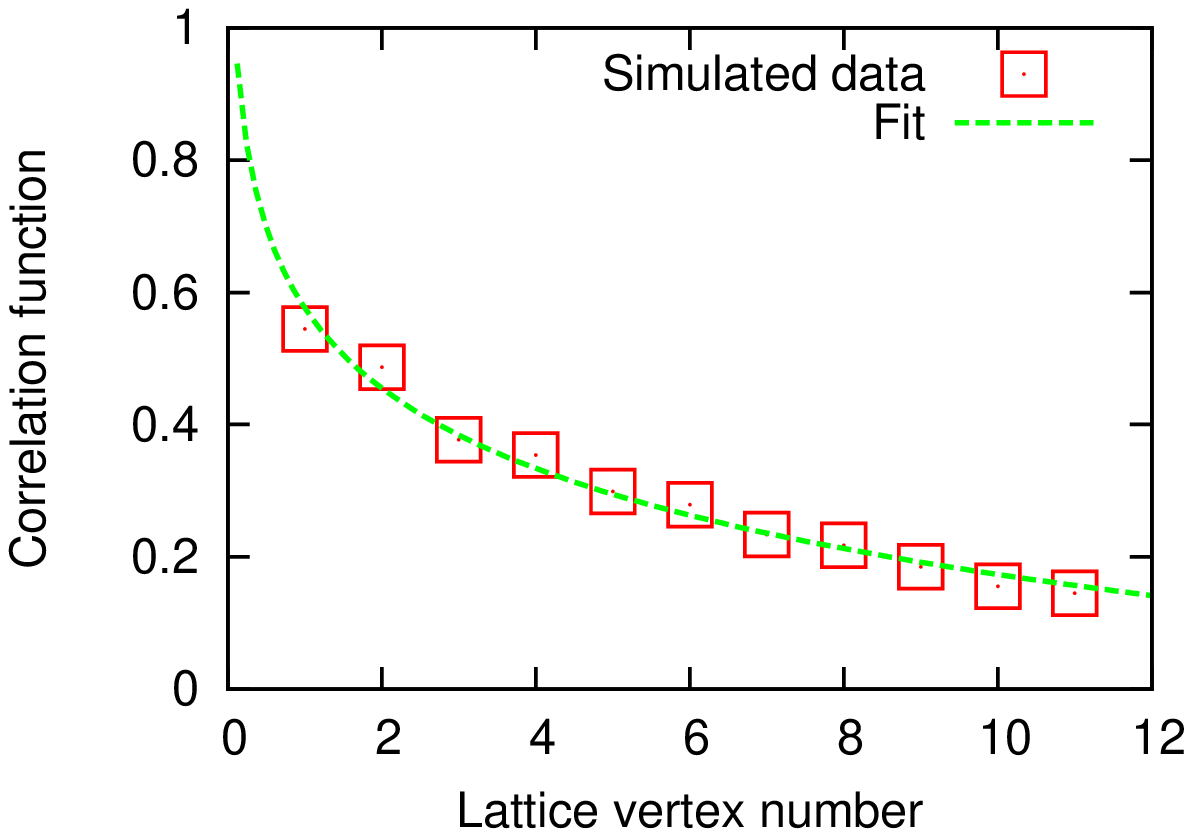}
  \includegraphics[scale=0.43,angle=-90]{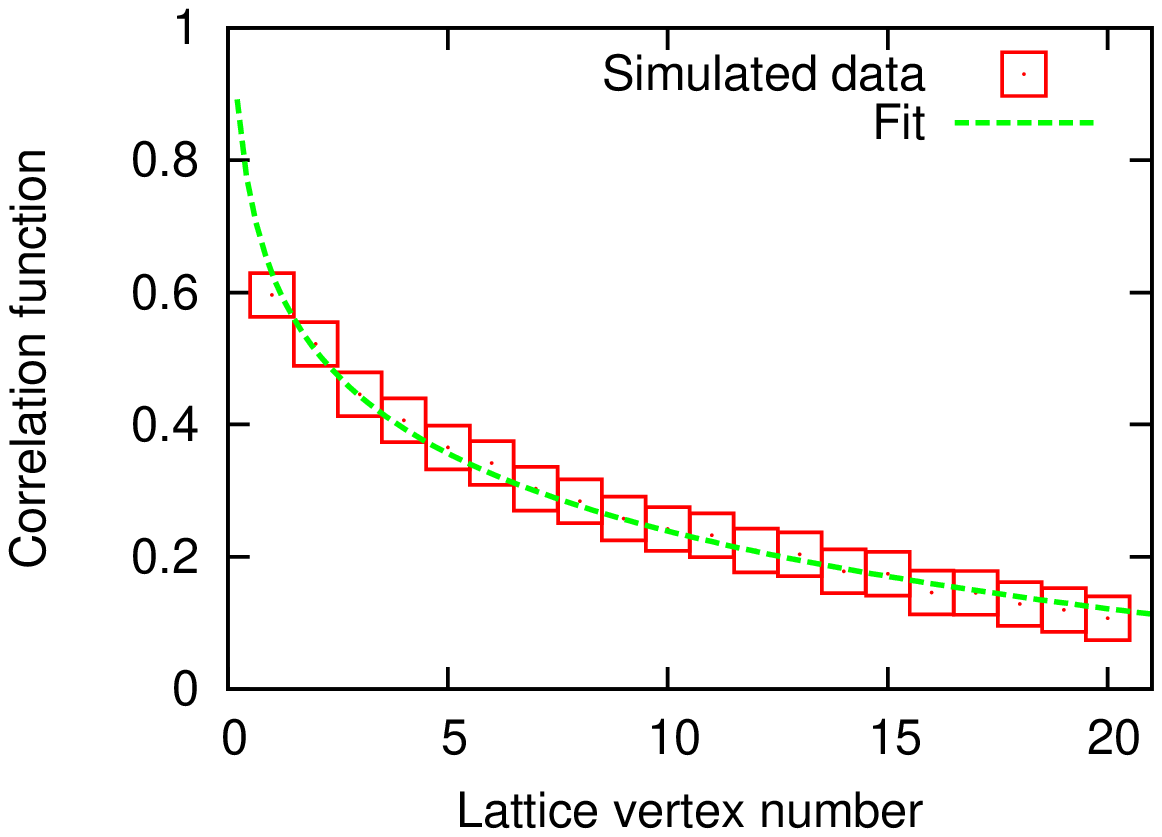}
  \includegraphics[scale=0.43,angle=-90]{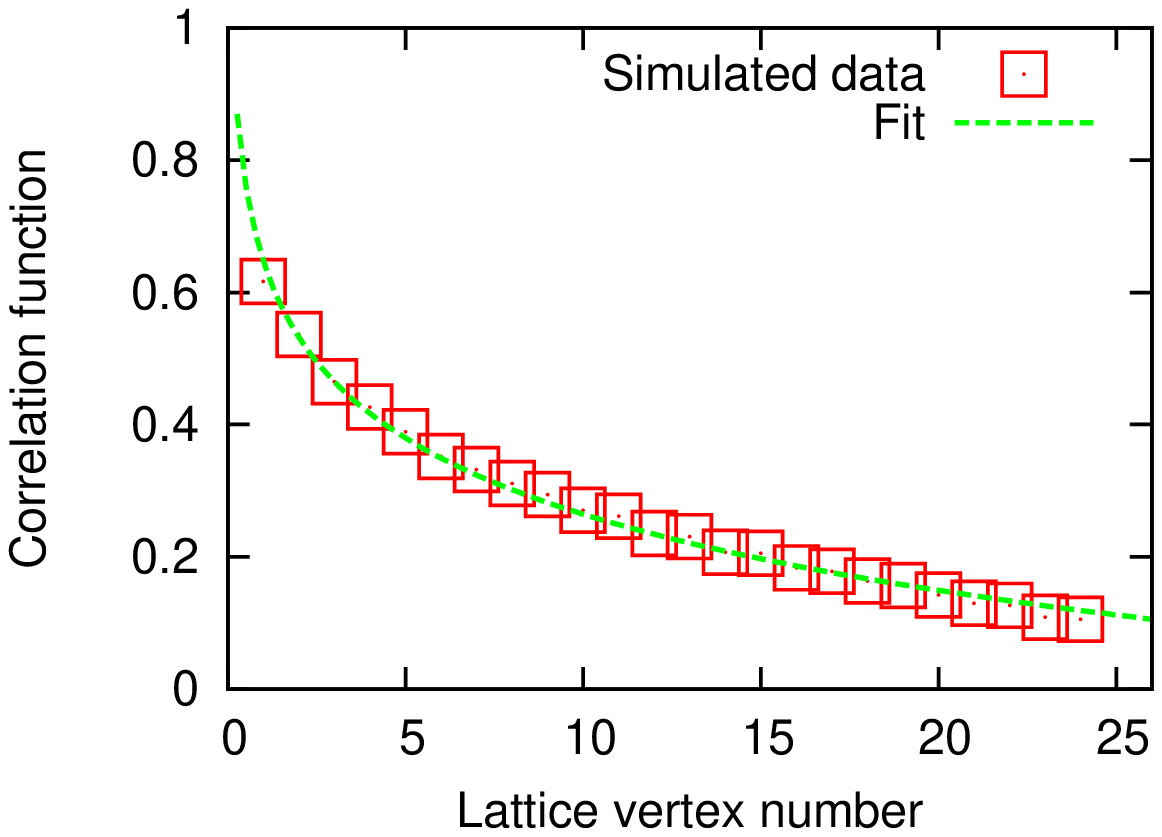}
  \caption{\label{fig:six-vertex-corr} A dependence of the two-point correlation function on a distance and a logarithmic fit for lattices of the size $30\times 30$ (left), $50\times 50$ (center), $60\times 60$ (right).}
\end{figure}

A logarithmic behavior, analogous to the dimer model, is observed for the six vertex model in case of $\Delta=0$.
Fitting the simulated data by the formula $a \log{|x-y|}+b$ with fixed $y$ and $a$,$b$ as the fitted parameters, as shown in Fig.~\ref{fig:six-vertex-corr},
we obtain the fitted values listed in Tab.~\ref{tttt1}. The normalization for the correlation function is the same, $\left<h(y)^{2}\right>=1$. 
%
\begin{table}[htpb]
\begin{center}
\begin{tabular}{c|ccccc}
    M&20&30&40&50&60\\
    \hline
    a&$-0.188 \pm 0.015$&$-0.175 \pm 0.008$&$-0.174 \pm 0.005$&$-0.169 \pm 0.004$&$-0.166 \pm 0.003$\\
    b& $0.528 \pm 0.024$&$0.576 \pm 0.014$&$0.609 \pm 0.011$&$0.629 \pm 0.008$&$0.647 \pm 0.007$
\end{tabular}
\caption{Values of the fitted parameters for the height-height correlation function in the six vertex model for $\Delta=0$.
}
\label{tttt1}
\end{center}
\end{table}
We see that the parameter $b$ has logarithmic dependence on $M$. Fitting $b(M)=-a\log M$, we obtain $a=-0.165\pm0.003$, which is in agreement with the values of $a$ in Tab.~\ref{tttt1}. Thus, we confirm a logarithmic behaviour of the correlation function inside the non-frozen region,
that is predicted by the Gaussian free field description.

\section{Conclusion}
\label{sec:further-work}

In this paper, we apply the Wang-Landau algorithm as well as the Metropolis algorithm to study the scaling behavior of the free energy in the dimer model. We compare a precision of the results obtained by these methods, as well as confront them to the theoretical prediction based on the combinatorial formula. We conclude that Wang-Landau algorithm gives more precise results and can be used for extracting the scaling parameters of the free energy for a domain with a complex geometry.
We also present the two-point correlation function and illustrate the limit shape phenomenon for the
dimer and six vertex model. A logarithmic dependence of the correlation function is in agreement with the
Gaussian free field description of fluctuations around the limit shape. 

\section*{Acknowledgments}
\label{sec:acknowledgements}
We are grateful to professor Nikolai Reshetikhin for valuable discussions and comments.
We are thankful to the organizers of the conference Physica.SPb/2018.
This research is supported by RFBR grant No. 18-01-00916.
The calculations were carried out using the facilities of the SPbU Resource Center ``Computational Center of SPbU''.

\section*{References}

\bibliographystyle{iopart-num}
\bibliography{listing,bibliography,dimers}{} 

\providecommand{\newblock}{}
\begin{thebibliography}{10}
\expandafter\ifx\csname url\endcsname\relax
  \def\url#1{{\tt #1}}\fi
\expandafter\ifx\csname urlprefix\endcsname\relax\def\urlprefix{URL }\fi
\providecommand{\eprint}[2][]{\url{#2}}

\bibitem{lieb1967exact}
Lieb E~H 1967 {\em Physical Review Letters\/} {\bf 18} 692

\bibitem{korepin1982}
Korepin V~E 1982 {\em Communications in Mathematical Physics\/} {\bf 86}
  391--418

\bibitem{tavares2015influence}
Tavares T, Ribeiro G and Korepin V 2015 {\em Journal of Physics A: Mathematical
  and Theoretical\/} {\bf 48} 454004

\bibitem{baxter2016exactly}
Baxter R~J 2016 {\em Exactly solved models in statistical mechanics\/}
  (Elsevier)

\bibitem{Fowler-1937}
Fowler R H;~Rushbrooke G~S 1937 {\em Transactions of the Faraday Society\/}
  {\bf 33} \urlprefix\url{https://doi.org/10.1039/TF9373301272}

\bibitem{doi:10.1080/14786436108243366}
Temperley H~N~V and Fisher M~E 1961 {\em The Philosophical Magazine: A Journal
  of Theoretical Experimental and Applied Physics\/} {\bf 6} 1061--1063
  (\textit{Preprint} \eprint{https://doi.org/10.1080/14786436108243366})
  \urlprefix\url{https://doi.org/10.1080/14786436108243366}

\bibitem{P.W-1961}
Kasteleyn P 1961 {\em Physica\/} {\bf 27}(12)
  \urlprefix\url{https://doi.org/10.1016/0031-8914(61)90063-5}

\bibitem{kenyon2009lectures}
Kenyon R 2009 {\em arXiv preprint arXiv:0910.3129\/}

\bibitem{kulish1990yang}
Kulish P, Reshetikhin N~Y and Sklyanin E 1990 Yang--baxter equation and
  representation theory: I {\em Yang-Baxter Equation In Integrable Systems\/}
  (World Scientific) pp 498--508

\bibitem{zj2000}
Zinn-Justin P 2000 {\em Phys. Rev. E\/} {\bf 62}(3) 3411--3418

\bibitem{ferrari}
Ferrari P~L and Spohn H 2006 {\em Journal of Physics A: Mathematical and
  General\/} {\bf 39} 10297

\bibitem{mangazeev}
Mangazeev V~V 2014 {\em Nuclear Physics B\/} {\bf 882} 70--96 ISSN 0550-3213

\bibitem{allison2005numerical}
Allison D and Reshetikhin N 2005 Numerical study of the 6-vertex model with
  domain wall boundary conditions {\em Annales de l'institut Fourier\/} vol~55
  pp 1847--1869

\bibitem{ks2018}
{Keating} D and {Sridhar} A 2018 {\em ArXiv e-prints\/} (\textit{Preprint}
  \eprint{1804.07250})

\bibitem{vershik1977kerov}
Vershik A~M and Kerov S~V 1977 Asymptotics of the plancherel measure of the
  symmetric group and the limiting form of young tableaux {\em Soviet Math.
  Dokl\/} vol~18 pp 527--531

\bibitem{kenyon2006dimers}
Kenyon R, Okounkov A and Sheffield S 2006 {\em Annals of mathematics\/}
  1019--1056

\bibitem{johansson2002non}
Johansson K 2002 {\em Probability theory and related fields\/} {\bf 123}
  225--280

\bibitem{pronko2010}
Colomo F, Pronko A~G and Zinn-Justin P 2010 {\em Journal of Statistical
  Mechanics: Theory and Experiment\/} {\bf 2010} L03002

\bibitem{tracy1994fredholm}
Tracy C~A and Widom H 1994 {\em Communications in mathematical physics\/} {\bf
  163} 33--72

\bibitem{metropolis1953equation}
Metropolis N, Rosenbluth A~W, Rosenbluth M~N, Teller A~H and Teller E 1953 {\em
  The journal of chemical physics\/} {\bf 21} 1087--1092

\bibitem{wang2001efficient}
Wang F and Landau D 2001 {\em Physical Review Letters\/} {\bf 86} 2050

\bibitem{landau2014guide}
Landau D~P and Binder K 2014 {\em A guide to Monte Carlo simulations in
  statistical physics\/} (Cambridge University Press)

\bibitem{kenyon2002laplacian}
Kenyon R 2002 {\em Inventiones mathematicae\/} {\bf 150} 409--439

\bibitem{vuletic2009generalization}
Vuleti{\'c} M 2009 {\em Transactions of the American Mathematical Society\/}
  {\bf 361} 2789--2804

\bibitem{reshetikhin2016limit}
Reshetikhin N and Sridhar A 2016 {\em arXiv preprint arXiv:1609.01756\/}

\bibitem{reshetikhin2017integrability}
Reshetikhin N and Sridhar A 2017 {\em Communications in Mathematical Physics\/}
  {\bf 356} 535--565

\bibitem{wolff1989collective}
Wolff U 1989 {\em Physical Review Letters\/} {\bf 62} 361

\bibitem{swendsen1987nonuniversal}
Swendsen R~H and Wang J~S 1987 {\em Physical Review Letters\/} {\bf 58} 86

\bibitem{bns}
Belov P~A, Nazarov A~A and Sorokin A~O 2017 {\em Phys. Rev. E\/} {\bf 95}(6)
  063308

\bibitem{bnsconf}
Belov P~A, Nazarov A~A and Sorokin A~O 2017 {\em Journal of Physics: Conference
  Series\/} {\bf 929} 012037

\bibitem{kenyon2000asymptotic}
Kenyon R 2000 {\em Acta Mathematica\/} {\bf 185} 239--286

\bibitem{sridhar2015asymptotic}
Sridhar A 2015 {\em arXiv preprint arXiv:1501.02057\/}

\bibitem{kenyon2001dominos}
Kenyon R 2001 {\em Annals of probability\/}  1128--1137

\bibitem{kenyon2008height}
Kenyon R 2008 {\em Communications in Mathematical Physics\/} {\bf 281} 675

\end{thebibliography}

\end{document}